\documentstyle[preprint,aps]{revtex}

\begin{document}

\title{Quantum Monte Carlo Calculations of $A\leq6$ Nuclei}
\author{B. S. Pudliner and V. R. Pandharipande}
\address{Physics Department, University of Illinois at Urbana-Champaign,
	 1110 West Green St., Urbana, Illinois 61801}
\author{J. Carlson}
\address{Theoretical Division, Los Alamos National Laboratory, Los Alamos
	 New Mexico 87545}
\author{R. B. Wiringa}
\address{Physics Division, Argonne National Laboratory, Argonne, IL 60439}

\date{January 11, 1995}
\maketitle

\begin{abstract}
The energies of $^{3}H$, $^{3}He$, and $^{4}He$ ground states, the
${\frac{3}{2}}^{-}$ and ${\frac{1}{2}}^{-}$ scattering states of $^{5}He$,
the ground states of $^{6}He$, $^{6}Li$, and $^{6}Be$ and the $3^{+}$ and
$0^{+}$ excited states of $^{6}Li$ have been accurately calculated
with the Green's function Monte Carlo method using
realistic models of two- and three-nucleon interactions.
The splitting of the $A=3$ isospin $T=\frac{1}{2}$ and $A=6$ isospin $T=1$,
$J^{\pi} = 0^{+}$ multiplets is also studied.
The observed energies and radii are generally well reproduced, however, some
definite differences between theory and experiment can be identified.
\end{abstract}
\narrowtext

A system of interacting nonrelativistic nucleons is the simplest model of
nuclei.  Even in this simple model exact calculations have been possible only
for a limited number of light nuclei due to the strong spin-isospin dependence
of nuclear forces.  For many years, only two-nucleon states could be exactly
calculated.  Next, the Faddeev method was used to study the three-nucleon
states \cite{Che85,Glo90}. In the past decade, many advances have
become possible due to the development of supercomputers.  Quantum Monte
Carlo methods were used to study nuclei with $A\leq5$ \cite{Car88,Car94},
the $^{4}He$ ground
state was calculated with the Faddeev-Yakubovosky method \cite{Glo93}, and
methods using hyperspherical functions were developed to study low-energy
three- and four-nucleon states \cite{Kie93}.  In this letter, we report the
first realistic six-nucleon (6N) quantum Monte Carlo calculations along
with updated results for nuclei with $A\leq5$.
Until now the $A=6$ nuclei have been mostly treated as three-body systems
with an $\alpha$ and two nucleons \cite{Sch93}.

The new Argonne $v_{18}$ two-nucleon interaction \cite{Wir95} is used here.
It is expressed as a sum of four parts:
\begin{equation}
   v = v_{14} + v_{cib} + v_{csb} + v_{em} .
\end{equation}
Its dominant part, $v_{14}$, contains 14 isoscalar operators as in the old
Argonne $v_{14}$ \cite{Wir84}.  The charge-independence-breaking part,
$v_{cib}$, has three isotensor terms with operators
$[3\tau_{iz}\tau_{jz} - {\bf \tau}_{i}\!\cdot\!{\bf \tau}_{j}]$ $\otimes$
$[1,{\bf \sigma}_{i}\!\cdot\!{\bf \sigma}_{j},S_{ij}]$, and
includes the effect of the mass difference between charged and neutral pions.
The $v_{cib}=0$ in isospin $T=0$ states, while in $T=1$ states
$v_{cib}(np)=-2v_{cib}(nn) =-2v_{cib}(pp)$.
The isovector charge-symmetry-breaking part, $v_{csb}$, contains the operator
$(\tau_{iz}+\tau_{jz})$; it accounts for the difference between $nn$ and $pp$
interactions and vanishes in $np$ pairs.
The electromagnetic part, $v_{em}$, contains $pp$ and $np$ Coulomb and magnetic
interactions in all pairs.  The kinetic energy operator associated with
this model has isoscalar and isovector parts denoted by $K$ and $K_{csb}$:
\begin{eqnarray}
K + K_{csb} = -\frac{\hbar^{2}}{4} \sum_{i} \left[\right. && (\frac{1}{m_{p}} +
\frac{1}{m_{n}}) \nabla^{2}_{i} \nonumber\\
&& + (\frac{1}{m_{p}} - \frac{1}{m_{n}})\tau_{iz}\nabla^{2}_{i} \left.\right]
\end{eqnarray}
Due to its careful treatment of isospin-symmetry-breaking terms, the new
Argonne $v_{18}$ model is well-suited to study the mass differences between the
$T=\frac{1}{2}$, $^{3}He - ^{3}H$ doublet and the
$T=1$, $J^{\pi} = 0^{+}$, $^{6}He - ^{6}Li - ^{6}Be$ triplet.

Three-nucleon interactions, $V_{ijk}$, described with the Urbana model
\cite{Joe83} are included in the nuclear Hamiltonian.
These contain a two-pion exchange part with its strength $A_{2\pi}$ chosen to
reproduce the observed binding energies of $^{3}H$ and $^{4}He$, and a
phenomenological spin-isospin independent interaction of strength $U_{o}$
adjusted to obtain the empirical equilibrium density of nuclear matter.
In Urbana models IX (VIII) of $V_{ijk}$, to be used in conjunction with the new
$v_{18}$ (old $v_{14}$), these parameters have values $A_{2\pi}=-0.0293$
$(-0.028)$ and $U_{o}=0.0048$ $(0.005)$ MeV.

The Green's function Monte Carlo (GFMC) calculations \cite{Car88,Car94}
are carried out with a simpler isoscalar Hamiltonian:
\begin{eqnarray}
\hat{H} = K + \sum_{i<j}v_{8}(ij) + \sum_{i<j<k}V_{ijk}.
\end{eqnarray}

The interaction $v_{8}(ij)$ has eight terms, with operators
$[1,{\bf \tau}_{i}\!\cdot\!{\bf \tau}_{j}]$ $\otimes$
$[1, {\bf \sigma}_{i}\!\cdot\!{\bf \sigma}_{j},S_{ij},{\bf L}\!\cdot\!{\bf
S}]$,
chosen such that it equals the $v_{14}$ in all $S$- and $P$-waves as well as in
the $^{3}D_{1}$ wave and its coupling to the $^{3}S_{1}$.
The eigenstates of $\hat{H}$ are computed, and all other terms, namely
$(v_{14}-v_{8}),v_{cib},v_{csb},v_{em}$, and $K_{csb}$,
are treated as first order perturbations.  The error in the binding energy of
$^{3}H$ due to the perturbative treatment of $v_{14}-v_{8}$ has been estimated
by Kamada and
Gl\"{o}ckle \cite{Glo94} to be $\sim0.02$ MeV.

Nuclear states are represented by vector functions $\Psi({\bf R})$ whose
components $\psi_{\alpha}({\bf R})$ give the amplitudes for the nucleons, at
spatial coordinates ${\bf R} = ({\bf r}_{1},{\bf r}_{2},\cdots,{\bf r}_{A})$,
to be in spin-isospin state $|\alpha\rangle$.
The GFMC calculation of the lowest energy state with quantum numbers
$\xi (=J^{\pi},T,T_{z})$ starts with an approximate wave function
$\Psi_{v}({\bf R})$ determined from a variational Monte Carlo (VMC)
calculation for that state.
It yields values of $\Psi(\tau,{\bf R}_{i}(\tau))$, where
\begin{eqnarray}
\Psi(\tau,{\bf R}) = e^{-\hat{H}\tau}\Psi_{v}({\bf R}),
\end{eqnarray}
at configurations ${\bf R}_{i}(\tau)$ distributed with probability
$|\Psi^{\dagger}_{v}({\bf R})\Psi(\tau,{\bf R})|$.
Here $i$ labels configurations which number $\sim10,000$.
The $e^{-\hat{H}\tau}$ is considered as a product of many small imaginary time
steps $e^{-\hat{H}\triangle\tau}$, with $\triangle\tau\sim0.0001 MeV^{-1}$.
Using the small-time limit of the propagator $e^{-\hat{H}\triangle\tau}$,
correct up to order $\triangle\tau$,
the $\Psi(\tau+\triangle\tau,{\bf R}_{i}(\tau+\triangle\tau))$ are
stochastically estimated from the known $\Psi(\tau,{\bf R}_{i}(\tau))$.

The number of spin-isospin states that contribute to the $T=0(1)$ 6N
$\Psi({\bf R})$ is $320(576)$, and their propagator
$e^{-\hat{H}\triangle\tau}$ is a $320\times320 (576\times576)$ complex
matrix function.
Therefore 6N GFMC calculations are numerically very intensive.
They were performed on the IBM SP parallel computer at Argonne National
Laboratory with 128 nodes operating at $\sim40$ MFLOPS/node.
Propagating $10,000$ 6N configurations up to $\tau=0.06 MeV^{-1}$ requires
$\sim2,000$ node hours.
In contrast, the $^{4}He$ ground state wave function $\Psi({\bf R})$ has
only $32$ spin-isospin states, and requires $\sim100$ node hours for a
similar calculation.

It is well known that there are no $A=5$ bound states.  The ${\frac{3}{2}}^{-}$
and ${\frac{1}{2}}^{-}$ resonances in $^{5}He$ have been studied as $n-^{4}He$
scattering and the phase shifts in these partial waves have been extracted
\cite{Arn73}.  The GFMC calculations for these states are carried out with an
external $\alpha-n$ potential that vanishes for
$|{\bf r}_{\alpha}-{\bf r}_{n}|<12.5$ fm and is infinitely repulsive for
$|{\bf r}_{\alpha}-{\bf r}_{n}|>12.5$ fm.
These conditions also imply that the external potential is 0($\infty$) for
$|{\bf r}_{n}|<10(>10)$ fm in the center of mass frame.
The calculated $\Psi_{o}$ corresponds to a scattering solution with a node
at $|{\bf r}_{\alpha}-{\bf r}_{n}|=12.5$ fm, however, only the interesting part
with $|{\bf r}_{\alpha}-{\bf r}_{n}|<12.5$ fm is retained in the calculation.
Assuming that the interaction between $\alpha$ and $n$ is negligible for
$|{\bf r}_{\alpha}-{\bf r}_{n}|>12.5$ fm the true energy of this
state can be obtained from the phase shifts \cite{Joe84}.

The variational wave functions used in these calculations include spatial and
spin-isospin two-body and three-body correlations denoted by
$f_{c}(r_{ij}), U_{ij},$ and $U_{ijk}$ in ref. \cite{Wir91} .  The uncorrelated
wave functions for $^{3}H$ and $^{4}He$ are as given in \cite{Wir91}, while
those for $^{5}He$ have an additional nucleon in the $(p{\frac{1}{2}})$
or $(p{\frac{3}{2}})$ scattering state.
The two extra nucleons in 6N states are in optimized superpositions of
$(p{\frac{3}{2}})^2$, $(p{\frac{1}{2}}p{\frac{3}{2}})$, and
$(p{\frac{1}{2}})^2$
states whose radii are constrained to reasonable values.
Details of the VMC and GFMC calculations will be published separately.

The transient energy $E(\tau)$ defined as:
\begin{eqnarray}
E(\tau) =
\frac{\langle\Psi_{v}|\hat{H}|\Psi(\tau)\rangle}{\langle\Psi_{v}|\Psi(\tau)\rangle}  = \frac{\langle\Psi(\frac{\tau}{2})|\hat{H}|\Psi(\frac{\tau}{2})\rangle}{\langle\Psi(\frac{\tau}{2})|\Psi(\frac{\tau}{2})\rangle} ,
\end{eqnarray}
provides an upper bound which approaches the lowest-energy eigenvalue with
quantum numbers $\xi$ in the limit $\tau\rightarrow\infty$.
The values of $E(\tau)$, calculated from $10,000$ configurations for each of
the
three 6N states and $20,000$ configurations for $^4He$
are shown in Fig.~\ref{fig1} along with their statistical
error estimates.  The $^{4}He$ $E(\tau)$ decreases from the variational
energy $E_{v} = E(\tau = 0)$ by $\sim1.2$ MeV in $\sim0.02 MeV^{-1}$, and is
almost independent of $\tau$ thereafter, suggesting that the $\Psi_{v}$ has
$\leq2\%$ admixture of states with excitation energy $\geq50$ MeV.
The $E(\tau)$ of the 6N states, in contrast, decreases by $\sim4$ MeV from
$E_{v}$, and does not appear to have reached the $\tau\rightarrow\infty$
asymptotic value at $\tau=0.06 MeV^{-1}$.  The statistical error, governed by
variance of $\hat{H}\Psi_{v}({\bf R})/\Psi_{v}({\bf R})$ is larger in the 6N
$E(\tau)$.
These indicate that the 6N $\Psi_{v}$ are not as accurate as the $^{4}He$
$\Psi_{v}$, and that they need to be improved for more accurate GFMC
calculations.

The average value of $E(\tau)$ in the $\tau$ interval $0.03$ to $0.06$
$MeV^{-1}$ is listed as $\bar{E}$ in Table I.
The $\bar{E}$ of $^{3}H$, $^{4}He$, and $^{5}He$ can be indentified with their
ground state energies, however,
those of the 6N states can only be regarded as upper bounds.  It is difficult
to extrapolate the 6N $E(\tau)$ to $\tau\rightarrow\infty$, particularly due to
the large statistical errors; nevertheless we have attempted it in the
following way.
Let $\Psi_{i}$ be the eigenstates with quantum numbers $\xi$ and energies
$E_{i}$.
The $\Psi_{v}$ contains admixtures of $\Psi_{i}$ with amplitude $\beta_{i}$
in addition to the $\Psi_{o}$ of the lowest-energy state.
Admixtures with the smallest $E_{i}-E_{o}$ determine the behavior of $E(\tau)$
at large $\tau$.
We approximate the contribution of low-lying states with two delta functions
at $E_{1}$ and $E_{2}$ with amplitudes $\beta_{1}$ and $\beta_{2}$.
It is experimentally known that the lowest $T=0$, $J^{\pi}=1^{+} (3^{+})$
resonances in $^{6}Li$ are $5.65 (\sim13.6)$ MeV above the $1^{+} (3^{+})$
bound states \cite{Ajz84}.
Accordingly, we assume that $E_{1}-E_{o} = 5.65 (13.6)$ MeV for the
$1^{+} (3^{+})$ states and also use the value $13.6$ for $T=1$,
$J^{\pi}=0^{+}$.
Further assuming that $E_{2}-E_{o} = 30$ MeV the calculated values of
$E(\tau>0.01 MeV^{-1})$ are fitted by varying $E_{o}$, $\beta_{1}$, and
$\beta_{2}$.  Fortunately, the fits are not very sensitive to $E_{2}-E_{o}$,
and are shown in Fig.~\ref{fig1}.  The resulting value of $E_{o}$ is listed
as the calculated energy in Table I.  The error in the
extrapolated $E_{o}$ is much larger than that in $\bar{E}$ and less reliable.

The expectation values of various terms in the Hamiltonian are also listed in
table I.  These are averages over the interval $\tau = 0.03$ to
$0.06 MeV^{-1}$, calculated using
\begin{eqnarray}
\langle\hat{O}(\tau)\rangle =
2\frac{\langle\Psi_{v}|\hat{O}|\Psi(\tau)\rangle}{\langle\Psi_{v}|\Psi(\tau)\rangle} - \frac{\langle\Psi_{v}|\hat{O}|\Psi_{v}\rangle}{\langle\Psi_{v}|\Psi_{v}\rangle} ,
\end{eqnarray}
correct up to order $|\Psi(\tau)\rangle - |\Psi_{v}\rangle$.
The dominant $\langle v_{14}\rangle$ and $\langle K\rangle$ have similar
values in $^{4}He$ and $^{5}He$; while in the 6N states they are closer to
the sum of their values in $^{4}He$ and $^{2}H$.
The  $\langle V_{ijk}\rangle$ are similar in $^{4}He$,
$^{5}He$, and 6N states.  The $\langle v_{LS}\rangle$ contains both
${\bf L}\!\cdot\!{\bf S}$ and
${\bf L}\!\cdot\!{\bf S} ({\bf \tau}_{i}\!\cdot\!{\bf \tau}_{j})$
terms in the NN interaction.  They contribute, along with $V_{ijk}$
\cite{Pie93}, to the splitting between ${\frac{3}{2}}^{-}$ and
${\frac{1}{2}}^{-}$ states of $^{5}He$.  However, the calculated splitting
of $0.8(3)$ MeV is much smaller than the observed $1.4$ MeV.  The magnitudes
of $\langle v_{LS}\rangle$ and $\langle V_{ijk}\rangle$ are larger in the
6N $0^{+}$ and $3^{+}$ states than in the $1^{+}$, suggesting that the
$1^{+}$ has less contribution from the $(p{\frac{3}{2}})^2$ configuration
than the other two.  The underbinding of the 6N $0^{+}$ and $3^{+}$ states by
$\sim1$ MeV is probably related to that of the $^{5}He$ ${\frac{3}{2}}^{-}$
state.  In $^{5}He$, only the expectation value $\langle v_{coul}^{pp}\rangle$
of the Coulomb interaction between protons is calculated.
The other terms in $\langle v_{em}\rangle$ for $^{5}He$ (given in parentheses
in table I) are estimated from results for $^{4}He$ and $^{6}He$.
The $\langle v_{coul}^{pp}\rangle$ decreases by $\sim5\%$ from $^{4}He$ to
$^{6}He$ indicating that the $\alpha$-cluster expands as we go from
$A=4$ to $6$.

The last three lines of Table I give rms proton and neutron radii.
The calculated values of $R(p)$ compare well with the those extracted from
observed charge radii \cite{DeV87}.  The experimental  $R(p)$ of $ ^{3}He$ is
$1.77$ fm, in reasonable agreement with the calculated $R(n)$ of $^{3}H$.
The Coulomb interaction accounts for most of the isovector $^{6}Be - ^{6}He$
difference (Table II).  Since this difference is correctly predicted, the
$R(p)$ of $^{6}Be$, assumed to be equal to the $R(n)$ of $^{6}He$, appears to
have a reasonable value.

The contributions of  $v_{em},v_{cib},v_{csb}$, and $K_{csb}$, treated as first
order perturbations in this work, are responsible for the energy differences
within the $T=\frac{1}{2}$ and $T=1$, $J^{\pi}=0^{+}$ 6N multiplets listed in
Table II.
The present Hamiltonian explains the isovector energy differences
$^{3}He - ^{3}H$ and $^{6}Be - ^{6}He$ fairly well.  The three-body
calculations show that the isovector $v_{csb}$ and $K_{csb}$ are necessary
to obtain the observed $^{3}He - ^{3}H$ difference, in agreement with earlier
results of Faddeev calculations \cite{Wu90}.  Unfortunately, the
calculated value of the isotensor difference
$\case{1}{2}(^{6}Be + ^{6}He) - ^{6}Li$ is much larger than observed.
The observed difference is essentially explained by the electromagnetic
interaction alone.  This is very puzzling because most
of the contribution of the isotensor $v_{cib}$ to this difference should be
from the relative $^{1}S_{0}$ two-nucleon state in which the difference
between $pp$ and $np$ phase shifts seems to be well established
\cite{Ber90,Sto93}.
The nonperturbative contribution of the Coulomb interaction, particularly in
$^{6}Be$, neglected here, may reduce the value of this isotensor difference.
There could also be some contribution from charge-dependence of the two-pion
exchange $V_{ijk}$.

In conclusion,  we have demonstrated that the GFMC method can be used to
accurately calculate the energies of the many nuclear states with $A\leq6$
from realistic models of nuclear forces.
It appears possible to extend these calculations to several $A=7$ and $8$
states by using larger parallel computers such as the IBM SP2.
The calculated energies are in good agreement with experiment.  However, some
differences, such as the underestimation of the splitting between
${\frac{3}{2}}^{-}$ and ${\frac{1}{2}}^{-}$ states of $^{5}He$ are clearly
established.  We could attempt to probe relativistic effects
\cite{Bro87,Car93},
and the spin-isospin dependence of the short-range part of the $V_{ijk}$
using these differences.  A detailed analysis of the GFMC wave function
$e^{-\hat{H}\tau}\Psi_{v}$ is in progress to study the
structure of the 6N states and improve upon their $\Psi_{v}$.

\acknowledgements
The 6N calculations were made possible by a generous grant of computer
time from the Mathematics and Computer Science Division of Argonne National
Laboratory.  We also received valuable support from the National Energy
Research Supercomputer Center at Livermore and the National Center for
Supercomputing Applications at Urbana.
The authors thank Dr. Steven C. Pieper for many useful suggestions.
The work of BSP and VRP is supported by the U.S. National Science Foundation
via grant PHY89-21025, that of JC by the U.S. Department of Energy, and that
of RBW by the U.S.  Department of Energy, Nuclear Physics Division, under
contract No. W-31-109-ENG-38.

\pagebreak

\begin{figure}
\caption{The top to bottom data sets show the $E(\tau)$ for $^6He$,
$^6Li(3^+)$,
$^4He$, and $^6Li(1^+)$ states, along with the fits described
in the text.  Constants of 2.5, 1.5, and 0.5 MeV have been added to the
$E(\tau)$ of $^6He$, $^6Li(3^+)$, and $^4He$, respectively, for clarity.}
\label{fig1}
\end{figure}

\pagebreak
\widetext
\begin{table}
\caption{Calculated energies and radii in MeV and fm.}
\begin{tabular}{llllllllll}
 Nucleus (J) & $^{2}H(1)$ &$^{3}H(\frac{1}{2})$ & $^{4}He(0)$
&$^{5}He(\frac{3}{2})$ & $^{5}He(\frac{1}{2})$& $^{6}He(0)$ &  $^{6}Li(1)$ &
$^{6}Li(3)$ \\
\tableline
E(Expt.)	 	&   --2.22&   --8.48&   --28.3&   --27.2&   --25.8&   --29.3&
--32.0&   --29.8\\
E(Calc.) 		&
--2.22&--8.47(2)&--28.3(1)&--26.5(2)&--25.7(2)&--28.2(8)&--32.4(9)&--28.9(6)\\
$\bar{E}$ 		&
--2.22&--8.47(2)&--28.3(1)&--26.5(2)&--25.7(2)&--27.3(4)&--31.1(4)&--28.2(3)\\
$\langle K\rangle$	&     19.9&   50.(1)&  118.(1)&  122.(2)&  117.(2)&
146.(4)&  143.(3)&  138.(3)\\
$\langle v_{14}\rangle$	&   --22.1&
--58.(1)&--142.(1)&--145.(2)&--140.(2)&--172.(4)&--173.(3)&--165.(3)\\
$\langle V_{ijk}\rangle$&      0. &--1.20(3)& --6.5(3)& --7.0(4)& --6.4(4)&
--7.0(7)& --6.2(6)& --6.9(5)\\
$\langle v_{LS}\rangle$	&   --0.08&--0.20(5)& --0.4(1)& --1.2(1)& --0.4(1)&
--2.7(3)& --1.5(5)& --3.0(4)\\
$\langle v_{em}\rangle$	&    0.018& 0.039(1)& 0.879(5)&   (0.86)&   (0.86)&
0.87(1)&  1.71(2)&  1.72(2)\\
$\langle v_{coul}^{pp}\rangle$	&      0. &      0. & 0.761(2)& 0.745(3)&
0.751(3)& 0.724(8)&  1.55(2)&  1.57(2)\\
R(n)			&    1.967&     1.72&  1.42(1)&  3.02(3)&  3.57(3)&  2.62(1)&  2.41(5)&
2.46(7)\\
R(p)			&    1.967&     1.58&  1.42(1)&  1.84(2)&  1.99(2)&  1.89(6)&  2.41(5)&
2.46(7)\\
R(p)(Expt.)		&    1.953&     1.61& 1.47    &         &         &         &
2.43   &         \\
\end{tabular}
\end{table}

\narrowtext
\begin{table}
\caption{Energy differences in MeV within isospin multiplets}
\begin{tabular}{lddd}
&$^{3}He-^{3}H$&$^{6}Be-^{6}He$&$\case{1}{2}(^{6}Be + ^{6}He) - ^{6}Li$\\
\tableline
$\langle v_{em}\rangle$		&   0.669&    2.33&   0.33\\
$\langle K_{csb}\rangle$	&   0.014&   0.036&   0.  \\
$\langle v_{csb}\rangle$	&   0.066&   0.116&   0.  \\
$\langle v_{cib}\rangle$	&   0.   &   0.   &   0.28\\
$\bigtriangleup(Calc.)$		&0.749(1)&  2.5(1)& 0.6(1)\\
$\bigtriangleup(Expt.)$		&   0.764&    2.35&   0.34\\
\end{tabular}
\end{table}
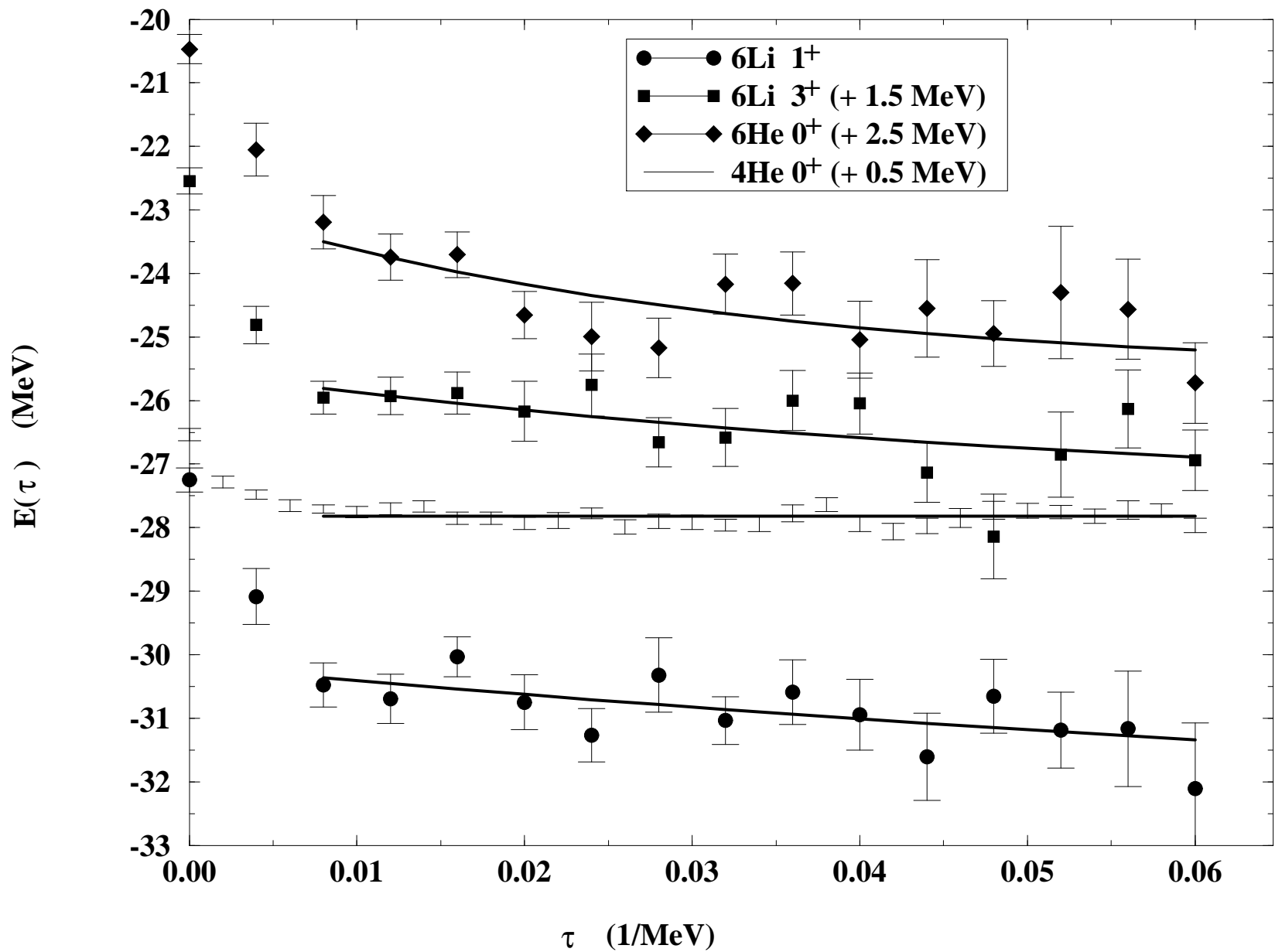
\end{document}